# Experimental Verification of the Spectral Shift between Near- and Far-Field Peak Intensities of Plasmonic Nanoantennas

*P. Alonso-González[1], P. Albella[1,2], F. Neubrech[3], Christian Huck[3], J. Chen[1,2], F. Golmar[1,4], F. Casanova[1,5], L. E. Hueso[1,5], A. Pucci[3], J. Aizpurua[2], and R. Hillenbrand[1,5]**

1  CIC nanoGUNE, 20018 Donostia – San Sebastián, Spain

2  Centro de Fisica de Materiales (CSIC-UPV/EHU) and Donostia International Physics Center (DIPC), 20018 Donostia-San Sebastián, Spain

3  Kirchhoff Institute for Physics, University of Heidelberg, Im Neuenheimer Feld 227, 69120 Heidelberg, Germany

4  I.N.T.I.–CONICET, Av. Gral. Paz 5445, Ed. 42, B1650JKA, San Martín, Bs As, Argentina

5  IKERBASQUE, Basque Foundation for Science, 48011 Bilbao, Spain

*r.hillenbrand@nanogune.eu

**ABSTRACT**

**Theory predicts a distinct spectral shift between the near- and far-field optical responses of plasmonic antennas. Here we combine near-field optical microscopy and far-field spectroscopy of individual infrared-resonant nanoantennas to verify experimentally this spectral shift. Numerical calculations corroborate our experimental results. We furthermore discuss the implications of this effect in surface-enhanced infrared spectroscopy (SEIRS).**

When a metal nanostructure is illuminated by light, the excitation of surface plasmons yields strongly concentrated optical fields at the metal surface, often referred as "hot spots"[1]. Metallic nanostructures can be thus considered as effective optical nanoantennas for converting propagating plane waves into localized fields[2]. This antenna function enables the control of electromagnetic fields at the nanometer scale[3] and thus has promoted the development of a vast variety of applications including surface-enhanced Raman scattering spectroscopy (SERS)[4-7], surface-enhanced infrared spectroscopy (SEIRS)[8,9], antenna-enhanced ultrafast nonlinear spectroscopy[10], near-field microscopy[11-13] and novel photo-detection schemes[14,15].

To explore new antenna functionalities or to optimize the antenna performance, different antenna designs have been developed[16-26]. A major goal is to achieve the highest local field enhancement, sensitivity and tunability of the antennas' optical response. These properties are essentially determined by the spectral position and width of the antenna resonance, which are typically studied by far-field spectroscopy[27]. However, surface-enhanced spectroscopies rely on light-matter interaction in the near field of the antenna, where an object is exposed to high field intensities in ultra-small volumes. Early theoretical studies[28] and recent publications[29-34] predict and indicate[35] that the spectral near- and far-field response of the antennas are shifted against each other, which might have implications for the application and optimization of optical antennas. Particularly, this spectral shift has been indicated in SERS[35] and antenna-mediated fluorescence[36] studies. However, such inelastic scattering experiments present several difficulties such as the inherent difference between excitation and scattered frequencies or the chemical bonding and charge transfer between molecules and metal nanostructures, preventing a rigorous experimental verification of the shift. Additionally, SERS experiments usually rely on measurements of samples exhibiting heterogeneously distributed hot spots of random field enhancement, which forces a statistical evaluation of the enhancement. In this work we circumvent these difficulties by measuring the local near fields of single antennas with a scattering-type near-field microscope, thus avoiding any inherent frequency shift between incident and scattered light or any chemical bonding or charge transfer between the probe and the antenna. Particularly, we study the fundamental dipolar mode of antennas resonant in the mid-infrared spectral range and subsequently discuss implications for surface- and antenna-enhanced infrared spectroscopy (SEIRS).

The spectral shift between near-field intensity $I_{NF}$ and far-field extinction $I_{ext}$ at the fundamental plasmon resonance is demonstrated in Fig. 1 for linear dipole antennas. We show numerical (finite- difference in time-domain FDTD, Lumerical Solutions) calcualtions of the near- and far-field optical response of 40 nm high and 140 nm wide Au rods of varying length L on a $CaF_2$ substrate. In Fig. 1a we display the spectral positions of the near-field peak intensity and the far-field peak extinction for different antenna lengths L. They were obtained by calculating $I_{NF}$ at the antenna extremity (marked by the red cross in the inset of Fig. 1a) and $I_{ext} = I_{in} - I_{trans}$, respectively, as a function of the illumination wavelength λ (Fig. 1a). $I_{in}$ the incident intensity and $I_{trans}$ is the calculated transmitted intensity in the far field of the antennas. The polarization of the incident light was parallel to the antenna axes. A spectral shift between the near-field and far-field peak intensities is observed throughout the whole spectral range from visible to mid-infrared wavelengths. At a fixed antenna length L, the near-field peak is shifted to a longer wavelength compared to the far-field extinction maximum. This shift is shown in more detail in Fig. 1b, where we plot $I_{ext}$ and $I_{NF}$ along the vertical dashed line in Fig. 1a (antenna length L=3.1 μm). Considering a fixed illumination wavelength λ, the near-field peak appears at smaller L compared to the far-field extinction. We illustrate this effect in Fig. 1c by plotting $I_{ext}$ and $I_{NF}$ along the horizontal line in Fig. 1a (λ=9.3 μm).

Based on Mie theory for small spheres, the shift between near- and far-field peak intensities was already studied by Messinger *et. al.* in the 80′s[28], however, only recently this phenomenon has been intuitively explained by describing metallic antennas as classically driven harmonic oscillators[32, 33]. When the harmonic oscillator is damped (which can be associated to dissipation in the antenna and to scattering losses), the maximum oscillation amplitude (which can be associated with the near-field amplitude) appears at lower frequencies than the maximum dissipation (which can be associated with the far-field absorption). While the spectral position of the maximum dissipation does not depend on the damping, the maximum oscillation amplitude generally shifts to lower energies with increasing damping. This explains why spectral shifts between near- and far-field peak intensities can be quite large in the case of strong plasmon damping, for example in plasmonic Ni antennas[33], or in strongly scattering antennas[37].

Here we experimentally verify the shift between near- and far-field peak intensities for mid-infrared antennas. To that end, we measure at a fixed wavelength both the near-field intensity (Fig. 2a) and far-field extinction (Fig. 2b) of individual antennas of varying length L, thus tracing the resonances according to Fig. 1c. This approach is used because quantitative near-field data are readily obtained by recording a single near-field image (Fig. 3a) of an antenna set where L is systematically varied[38, 39]. Furthermore, in applications such as in SEIRS, the antenna length is the essential parameter to be matched to the fixed vibrational resonance of the molecules under study[9, 40].

The antennas are Au rods of 40 nm height, 140 nm width and a length varying from L=2 μm to L=4.4 μm, fabricated by electron beam lithography on a $CaF_2$ substrate. The distance between the different antennas is 10 μm, which allows for measuring a far-field extinction spectrum of each individual antenna, as well as for recording a single near-field image of the whole antenna set (Fig. 3a). We thus obtain for each individual antenna both near-field intensity and far-field extinction, allowing for a quantitative comparison of the two quantities as a function of the antenna length L.

Near-field imaging (Fig. 2a) is performed with a side-illumination scattering-type scanning near-field optical microscope (s-SNOM, from Neaspec GmbH). A Si tip oscillating vertically at frequency $\Omega$ is used for locally scattering the antenna near-fields[21, 33, 38, 39, 41-43]. Both tip and antenna are illuminated with s-polarized infrared light from a $CO_2$ laser at an angle of 50º from the surface normal. The light scattered by the tip is recorded with a pseudo-heterodyne interferometer[44]. By locating a polarizer in front of the detector, we select the horizontally polarized scattered light. Demodulation of the detector signal at a higher harmonic frequency $n\Omega$ yields background free near-field signals[21, 42]. We note that by (*i*) illuminating the antennas with s-polarized light and (*ii*) detecting the s-polarized backscattered light, the demodulated amplitude signal $s_n$ yields the square of the local near-field amplitude[39],

$I_{NF} = s_n$.

Far-field spectroscopy (Fig. 2b) is performed with an infrared microscope (Bruker Hyperion 1000) coupled to a Fourier transform spectrometer (Bruker Tensor 27), yielding infrared extinction spectra of the same set of individual antennas. The antennas were illuminated with thermal radiation (polarization parallel to the long axis of the antenna) of intensity $I_{in}$ under normal incidence. To address individual antennas, the illumination is through a 10x10 μm size aperture. The transmitted light $I_{trans}$ is recorded with a resolution of 8 cm$^{-1}$, yielding far-field extinction spectra $I_{ext}= I_{in} - I_{trans}$. These were normalized to the extinction measured at least 30 μm away from the antennas (reference spectrum).

Fig. 3a shows the near-field image of the antenna set, revealing the typical dipolar mode pattern for each antenna (two bright spots, indicating the strongly concentrated near fields at the rod extremities)[41]. The near-field signal increases with increasing antenna length, until it reaches its maximum at L≈3.1 μm (marked by position C). With further increasing antenna length, the near-field signal decreases. This observation clearly reveals the resonance behavior of the antennas[45]. In Fig. 3b we show the far-field extinction spectra of the individual antennas marked in Fig. 3a. We observe how the far-field resonance (extinction maximum) shifts to longer wavelengths when the antenna length increases, following the typical behavior of dipole antennas[9].

To compare the near-field (NF) and far-field (FF) optical responses of the antennas, we plot in Fig. 4 both $I_{NF}$ and $I_{ext}$ as a function of the antenna length L for the fixed illumination wavelength λ=9.3 μm. The near-field intensities have been extracted from Fig. 3a at the extremities of the antennas where the maximum value is obtained (the cross in Fig. 3a marks the typical position). We point out that we study the fundamental antenna mode, which implies that the near-field spectrum is the same at every point on the surface of the antenna. The extinction has been extracted from the individual far-field extinction spectra of the antennas, some of them shown in Fig. 3b. We clearly see that the near-field and far-field peak intensities are shifted against each other. The far-field extinction maximum occurs at L=3.30 μm (marked by the black dashed line in Fig. 4), while the near-field maximum appears at a shorter antenna length of L=3.15 μm (marked by the red dashed line in Fig. 4). These experimental results indeed verify the shift between the calculated near-field intensity spectrum (red line in Fig. 4, obtained at the antenna extremity as done in Fig. 1) and the calculated far-field extinction spectrum $I_{ext} = I_{in} - I_{trans}$ (black curve in Fig. 4, obtained in the far-field of the antennas as introduced in Fig. 1) of the antennas. We note that no fitting of the numerical results has been applied, just normalization of all the curves to their maximum values. We also note that the use of dielectric tips together with the s-polarized sample illumination has been shown to faithfully measure the spectral near-field response of plasmonic antennas without introducing spectral shifts[46-49]. We thus can exclude that the significant spectral shift between the near- and far-field response of the antennas is introduced by the tip.

The shift between near- and far-field peak intensities may have important implications for sensing applications, as the optical interaction between molecules and antennas is mediated by the near field. The spectroscopic information about the molecules, however, is measured in the far field. In order to elucidate the influence of the shift in antenna-enhanced infrared extinction spectroscopy, we performed a numerical study of the near- and far-field response of molecules in the vicinity of infrared antennas. We consider Au antennas covered with a 20-nm-thick layer of polymethyl methacrylate (PMMA) molecules on a $CaF_2$ substrate. PMMA is used as an example because of its well-defined infrared vibrational resonance. Fig. 5a shows the calculated extinction $I_{ext}^{PMMA}$ of PMMA-coated antennas of different lengths L (black curves). With increasing L, the antenna resonance shifts to longer wavelengths, whereas the vibrational response of PMMA appears fixed at λ=5.8 μm, independently of L. Importantly, the spectral PMMA response is enhanced when it is close to the antenna resonance. Simultaneously, its line-shape is modified, which results from the Fano-like interference of both infrared resonances[9, 40, 50-52]. To isolate and quantify the antenna-enhanced spectral response of PMMA, we show in Fig. 5b the difference spectra $C = I_{ext}^{PMMA} - I_{ext}^*$, where $I_{ext}^*$ (green curves in Fig. 5a) is the antenna resonance in absence of the vibrational response of PMMA (baseline subtraction). We now define the "fingerprint contrast" ΔC as the difference between the maximum and the minimum values of each individual spectrum C (illustrated by the schematics in Fig.

5b). In Fig. 5c we show ΔC for the different antenna lengths L. We find that the largest "fingerprint contrast" is obtained for L=1.7 μm. By comparing ΔC with the near-field intensity and far-field extinction (both shown in Fig. 5d), we interestingly observe that the largest fingerprint contrast is obtained when the near-field intensity $I_{NF}^{PMMA}$ (rather than the far-field extinction $I_{ext}^{PMMA}$) reaches its maximum. This finding indicates that the spectral shift between near- and far-field resonances has indeed to be considered when optimizing antennas for spectroscopy applications.

In Fig. 5e we compare the near- and far-field spectra of the 1.7-μm-long PMMA-coated antennas (black solid and red dashed curves, respectively), where the fingerprint contrast in the far-field extinction spectrum is largest. We observe that the near-field intensity peak of the antenna matches the molecular vibration, rather than the far-field extinction maximum. To see this more clearly, we show the near-field intensity $I_{NF}^{*}$ (red solid line) and far-field extinction $I_{ext}^{*}$ (green solid line) in absence of the vibrational response of PMMA. In addition to the spectral shift, we observe that the spectral shape of the molecular fingerprint also differs in the near-field and far-field signals. We note that further extended experimental and theoretical studies are needed for more detailed insights into this phenomenon, as well as to draw general conclusions regarding the optimal spectral contrast, since a variety of antenna designs, molecule vibrations and physical spectroscopy processes (extinction, Raman scattering, fluorescence) exists. Nevertheless, the canonical case of a dipolar antenna for SEIRS, as studied in this work, already stresses the importance and necessity of considering spectral shifts between near- and far-field peaks, in order to optimize surface- and antenna-enhanced spectroscopies.

In conclusion, having performed experimental studies of individual infrared-resonant antennas by near-field microscopy and far-field extinction spectroscopy, we confirm experimentally the spectral shift between the near- and far-field peak intensities. Furthermore, we have studied numerically the implications of this spectral shift in SEIRS, showing that it has to be considered in order to optimize the molecular spectral absorption contrast in plasmonic (bio)-sensing devices.

**FIGURE CAPTIONS**

**Figure 1**: Numerical study of the near-field intensity $I_{NF}$ and far-field extinction $I_{ext}$ of linear dipole Au antennas on a $CaF_2$ substrate. (a) Spectral positions of the near-field peak intensity and the far-field peak extinction. The red and black lines are guides to the eye. The inset illustrates where the near-field intensity and the far-field extinction were evaluated. (b) Near-field intensity $I_{NF}$ and far-field extinction $I_{ext}$ as a function of wavelength $\lambda$ for an antenna length L=3.1 μm (spectra along the vertical dashed line in Fig. 1a). (c) Near-field intensity $I_{NF}$ and far-field extinction $I_{ext}$ as a function of nanorod length L for a fixed illumination wavelength $\lambda$=9.3 μm (spectra along the solid horizontal line in Fig. 1a). The spectra in (b) and (c) were normalized to their maximum values.

**Figure 2**: Illustration of the detection schemes employed for measuring the near- and far-field response of individual infrared antennas. (a) s-SNOM. (b) Infrared micro-spectroscopy.

**Figure 3**: Experimental near- and far-field study of linear infrared dipole antennas. (a) s-SNOM image of the nanoantennas at $\lambda$=9.3 μm. From the top to the bottom and from the left to the right, the antenna length increases from L=2 μm to L=4.4 μm. The cross indicates the position where the near-field intensities displayed in Fig. 4 have been measured. (b) Far-field extinction spectra of the antennas marked in Fig. 3a by the letters A-E. The vertical dashed line indicates the wavelength $\lambda$=9.3 μm, where the extinction values displayed in Fig. 4 have been extracted.

**Figure 4**: Experimental and calculated near-field intensity $I_{NF}$ (red) and far-field extinction $I_{ext}$ (black) as a function of the antenna length. The experimental near-field intensities were measured at the extremity of the individual nanoantennas, at the position indicated by a cross in Fig. 3b. The experimental far-field extinction values were extracted from the far-field spectra at $\lambda$=9.3 μm, as indicated by a vertical

dashed line in Fig. 3b. The red solid line shows the numerically calculated near-field intensity at the rod extremity (as indicated by the inset in Fig. 1a). The black solid line shows the numerically calculated total extinction evaluated in the far field of the antennas. Both, experimental and calculated data were normalized to the corresponding maximum values.

**Figure 5**: Numerical study of antenna-enhanced extinction spectroscopy of polymethyl methacrylate (PMMA) molecules. (a) Extinction spectra of PMMA-coated Au antennas on a $CaF_2$ substrate for different lengths L ($I_{ext}^{PMMA}$, black curves). The green curves show the extinction spectra of the antennas in absence of vibrational response, $I_{ext}^{*}$ (see methods). (b) Difference spectra $C = I_{ext}^{PMMA} - I_{ext}^{*}$. (c) "Fingerprint contrast" $\Delta C$ as defined in the scheme of (b). (d) Near-field intensity ($I_{NF}^{PMMA}$, red curve) and far-field extinction ($I_{ext}^{PMMA}$, black curve) of PMMA-coated Au antennas as a function of antenna length L. The illumination wavelength is λ=5.8 μm, matching the vibrational resonance of the PMMA molecules. (e) Near-field intensity ($I_{NF}^{PMMA}$, red dashed curve) and far-field extinction ($I_{ext}^{PMMA}$, black solid curve) of PMMA-coated Au antennas as a function of illumination wavelength λ. For comparison, we also show the near-field intensity $I_{NF}^{*}$ (red solid line) and far-field extinction $I_{ext}^{*}$ (green solid line) in absence of the vibrational response of PMMA. The antenna length is L=1.7 μm, corresponding to the maximum of $\Delta C$ in Fig. 5c.

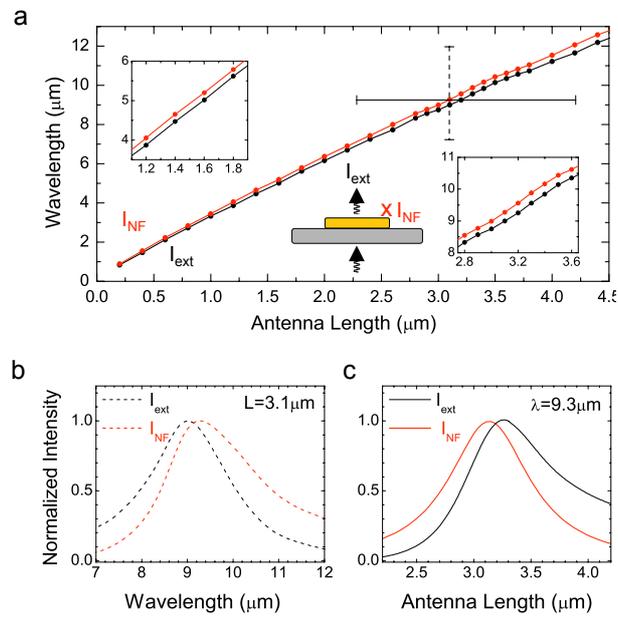

**Alonso-González et al., Figure 2**

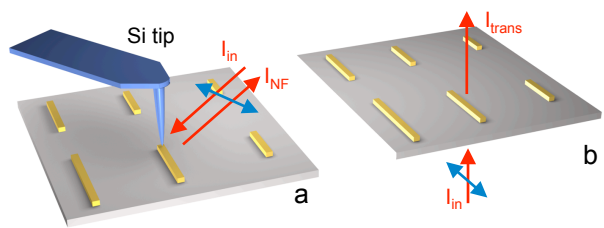

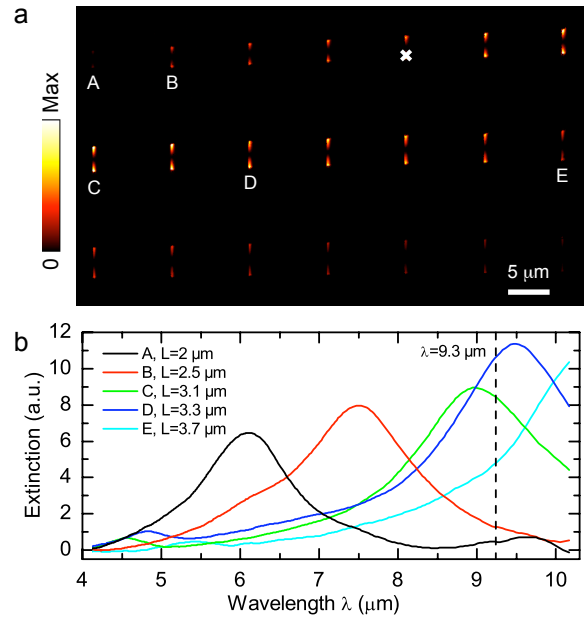

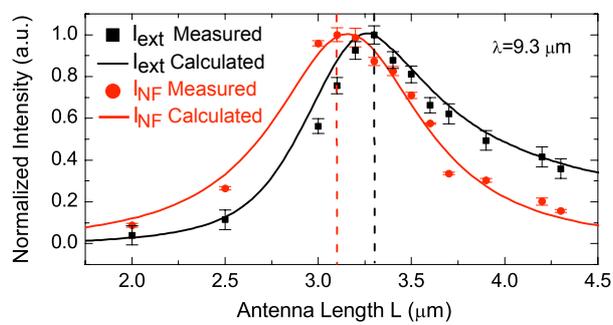

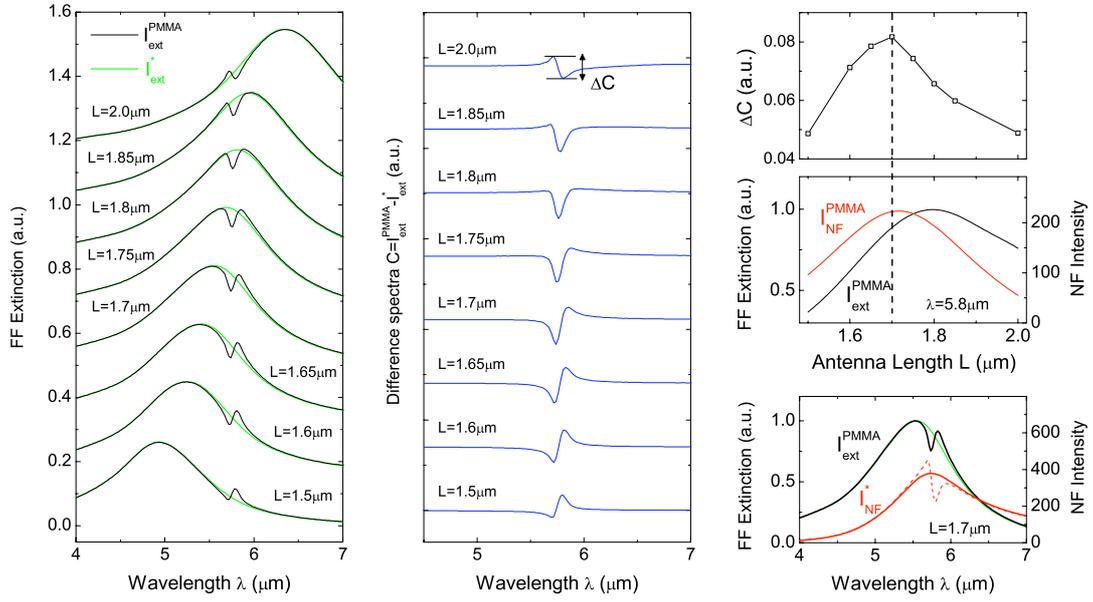